\documentstyle[12pt]{article}
\begin{document}

\title{Burgers velocity fields and dynamical transport processes}
 
\author{Piotr Garbaczewski and  Grzegorz Kondrat\\
Institute of Theoretical Physics, University of Wroc{\l}aw,
pl. M. Borna 9, \\
PL-50 204 Wroc{\l}aw, Poland}

\maketitle
\hspace*{1cm}
PACS numbers:  02.50-r, 05.20+j, 03.65-w, 47.27.-i

\begin{abstract}
We explore a connection of the  forced Burgers
equation with  the  Schr\"{o}dinger (diffusive) 
interpolating dynamics in the presence of deterministic
external forces.
This entails an exploration of the consistency conditions
that allow to interpret dispersion of passive contaminants in the
Burgers flow as a Markovian diffusion process. In general, the 
usage of a continuity equation 
$\partial _t\rho =-\nabla (\vec{v}\rho )$, 
where  $\vec{v}=\vec{v}(\vec{x},t)$ stands for the Burgers field and 
$\rho $ is the  density of transported matter, is at variance with the 
explicit diffusion scenario. 
Under these circumstances, we give a complete 
characterisation of the  diffusive matter transport that is 
governed by  Burgers velocity fields.  The result extends both to
the approximate description of the transport driven by an 
incompressible   fluid   and to motions in an infinitely 
compressible medium.

\end{abstract}

\newpage

The Burgers equation with (typically without, \cite{burg,hopf})
the forcing  term $\vec{F}(\vec{x},t)$:
$${\partial _t\vec{v} + (\vec{v}\nabla )\vec{v} = \nu 
\triangle \vec{v} +
\vec{F}(\vec{x},t)}\eqno (1)$$
and especially its statistically relevant $curl\, \vec{v}=0$
solutions (in below we shall not use  random initial data),
recently  have acquired  a considerable
popularity in  various physical contexts, \cite{zeld}-\cite{walsh}.

Although Burgers velocity fields  can be 
 analysed on their own, frequently one needs a supplementary 
insight into the matter transport dynamics that is
consistent with the  chosen (Burgers) velocity field evolution.
Then, the passive scalar (tracer or contaminant)  
advection-in-a-flow  problem, \cite{siggia,parisi,monin} naturally 
appears through the  parabolic dynamics:
$${\partial _tT + (\vec{v}\nabla )T = \nu \triangle T}\eqno (2)$$
While looking for the stochastic implementation of the microscopic 
(molecular) dynamics (2), \cite{woy2,parisi,monin},  it is  assumed
that the  "diffusing scalar" (contaminant  in the lore of early 
statistical turbulence models)  obeys an It\^{o} equation:  
$${d\vec{X}(t) =\vec{v}(\vec{x},t)dt  + \sqrt{2\nu } d\vec{W}(t)}
\eqno (3)$$
$$\vec{X}(0)=\vec{x}_0 \rightarrow \vec{X}(t)=\vec{x}$$
where the given  forced Burgers velocity field is    
perturbed by the noise term representing a molecular diffusion.
In the It\^{o} representation of diffusion-type random variable
$\vec{X}(t)$ one explicitly refers to the Wiener process  
$\sqrt {2\nu }\vec{W}(t)$, 
instead of the usually adopted formal white noise integral 
$\int_0^t \vec{\eta }(s)ds$, coming from the  Langevin-type version
of (3).

Under these premises, we cannot view equations (1)-(3) as 
completely independent (disjoint) problems:
the velocity field $\vec{v}$ cannot be  arbitrarily
inferred from  (1) or any other velocity-defining equation without
verifying the \it consistency \rm conditions, which would allow to
associate (2) and (3)  with 
a well defined random dynamics (stochastic process), and 
Markovian diffusion in particular, \cite{fried,horst}.

In connection with the usage of Burgers velocity fields (with
or without external forcing) which in (3) clearly are
intended to replace the 
standard \it forward  drift \rm of the would-be-involved Markov
diffusion process,  we have not found in the literature any 
attempt to resolve apparent contradictions arising if (2) and/or (3) 
are  defined by means of (1).
Also, an issue of the necessary \it correlation \rm (cf.
\cite{monin},
Chap.7.3, devoted to the turbulent transport and the related
dispersion of contaminants)  between the probabilistic
Fokker-Planck
dynamics of the diffusing tracer, and this of the  passive tracer 
(contaminant)  concentration (2),  has been  left aside in the
literature. 

Moreover, rather obvious hesitation could have been  observed
in attempts to establish the  most appropriate  matter transport 
rule,  if (1)-(3)  are adopted. 
Depending  on the particular phenomenological departure point,
 one either adopts the standard continuity equation,
 \cite{zeld,alb}, that is certainly
valid to a high degree of  accuracy in the low viscosity limit 
$\nu \downarrow 0$  of (1)-(3),  but incorrect on
mathematical grounds \it if \rm  there is a
diffusion involved \it and \rm simultaneously a solution of (1) stands 
for the respective \it current \rm velocity of the flow: 
${\partial _t\rho (\vec{x},t)= - \nabla 
[\vec{v}(\vec{x},t)\rho (\vec{x},t)]\enspace . }
$ 
Alternatively, following  the white noise calculus tradition telling
that  the stochastic integral
$\vec{X}(t)=\int_{0}^{t} \vec{v}(\vec{X}(s),s)ds +
\int_{0}^{t} \vec{\eta }(s)ds$
necessarily implies the Fokker-Planck equation, one adopts,
\cite{woy2}:
${\partial _t\rho (\vec{x},t) = \nu \triangle \rho (\vec{x},t) - 
\nabla [\vec{v}(\vec{x},t)\rho (\vec{x},t)]}$
which is clearly problematic in view of the classic Mc Kean's
discussion
of the  propagation of  chaos for the Burgers equation,
\cite{kean,cald,osa} and the derivation of the stochastic 
"Burgers process" in this  context: 
"the fun begins in trying to describe this Burgers motion as the 
path of a tagged molecule in an infinite bath of like molecules", 
\cite{kean}.

To put things on the solid ground, let us consider a Markovian
diffusion  process, which is characterised by the
transition probability density
(generally inhomogeneous in space and time law of random
displacements)
$p(\vec{y},s,\vec{x},t)\, ,\, 0\leq s<t\leq T$, and the probability 
density $\rho (\vec{x},t)$  of its random variable 
$\vec{X}(t)\, ,\,  0\leq t \leq T$. The process is completely 
determined  by these data. For clarity of discussion, we do not
impose  any spatial boundary  restrictions, nor fix any concrete 
limiting value of $T$ which, in principle, can be moved to infinity.

The conditions valid for any $\epsilon >0$: \\
(a) there holds
$lim_{t\downarrow s}{1\over
{t-s}}\int_{|\vec{y}-\vec{x}|>\epsilon }
p(\vec{y},s,\vec{x},t)d^3x=0$,\\
(b) there exists a (forward) drift
$\vec{b}(\vec{x},s)=lim_{t\downarrow s}{1\over {t-s}}\int_{|\vec{y}-
\vec{x}|
\leq \epsilon }(\vec{y}-\vec{x})p(\vec{x},s,\vec{y},t)d^3y$, \\
(c) there exists  a diffusion function (in our case  it is simply
a diffusion coefficient $\nu $)
$a(\vec{x},s)=lim_{t\downarrow s}{1\over {t-s}}
\int_{|\vec{y}-\vec{x}|\leq \epsilon }  (\vec{y}-\vec{x})^2
p(\vec{x},s,\vec{y},t)d^3y$,\\
are conventionally interpreted to define a diffusion process,
\cite{horst,fried}.
Under suitable restrictions (boundedness of involved
functions, their continuous differentiability) the function:
$${g(\vec{x},s)=
\int  p(\vec{x},s,\vec{y},T)g(\vec{y},T) d^3y }\eqno (4)$$
satisfies the backward diffusion equation (notice that the 
minus sign appears, in comparison  with (2))
$${- \partial _sg(\vec{x},s) = \nu \triangle g(\vec{x},s)  +
[\vec{b}(\vec{x},s)\nabla ]g(\vec{x},s)
\enspace .}\eqno (5)$$
Let us point out that the validity of (5) is known to be a \it
necessary
 \rm condition for the existence of a Markov diffusion process, whose
probability density $\rho (\vec{x},t)$ is to obey the Fokker-Planck 
equation (the forward drift $\vec{b}(\vec{x},t)$ replaces the
previously  utilized Burgers field $\vec{v}(\vec{x},t)$)): 
$${\partial _t\rho (\vec{x},t) = \nu \triangle 
\rho (\vec{x},t) - \nabla [\vec{b}(\vec{x},t)\rho (\vec{x},t)]}
\eqno (6)$$

The case  of particular interest in the  
nonequilibrium statistical physics literature  
appears when $p(\vec{y},s,\vec{x},t)$
is  a \it fundamental solution \rm of (5) with respect to variables
$\vec{y},s$,  \cite{krzyz,fried,horst},
see however \cite{olk2} for  an alternative situation.
Then, the transition probability density satisfies \it also \rm
the second Kolmogorov (e.g. the Fokker-Planck) equation in the
remaining $\vec{x}, t$ pair of variables.  Let us emphasize that
these two equations form an \it adjoint pair \rm , 
referring to the  slightly counterintuitive for
physicists, although transparent for
mathematicians, \cite{haus,fol,nel,zambr,zambr1}, issue of time
reversal of diffusions.

After  adjusting (3) to the present  context, $\vec{X}(t)=
\int_0^t\, \vec{b}(\vec{X}(s),s)\, ds + \sqrt {2\nu } \vec{W}(t) $
we can utilize  standard rules of the It\^{o} stochastic calculus,
\cite{nel1,nel,zambr,zambr1},  to realise that for any smooth function
$f(\vec{x},t)$ of the random variable $\vec{X}(t)$ the 
conditional expectation value:
$$lim_{\triangle t\downarrow 0} {1\over {\triangle t}}\bigl [\int
p(\vec{x},t,\vec{y},t+
\triangle
t)f(\vec{y},t+\triangle t)d^3y - f(\vec{x},t)\bigr ] = 
(D_+f)(\vec{X}(t),t)=    \eqno (7) $$
$$=  (\partial _t + (\vec{b}\nabla )+ \nu \triangle )f(\vec{x},t)$$
$$\vec{X}(t)=\vec{x}$$
determines  the forward drift $\vec{b}(\vec{x},t)$ (if we set 
components of $\vec{X}$ instead of $f$) and allows to introduce  
the local field of forward accelerations associated with the 
diffusion process, which we constrain by demanding (see e.g. Refs. 
\cite{nel1,nel,zambr,zambr1} for prototypes of such dynamical 
constraints):
$${(D^2_+)\vec{X})(t) =(D_+\vec{b})(\vec{X}(t),t) =(\partial _t
\vec{b} +
(\vec{b}\nabla )\vec{b} + \nu \triangle
\vec{b})(\vec{X}(t),t)= \vec{F}(\vec{X}(t),t)}\eqno (8)$$
where, at the moment arbitrary,  function $\vec{F}(\vec{x},t)$ 
may be interpreted as the external forcing applied to  
the diffusing system,  \cite{blanch}. 
In particular, if we assume that drifts remain
gradient fields, $curl \, \vec{b}= 0$, under the forcing, then
those that are allowed by the prescribed choice of
$\vec{F}(\vec{x},t)$    \it must \rm fulfill the compatibility
condition (notice the conspicuous absence of the standard
Newtonian minus sign in this analogue of the second Newton law)
$${\vec{F}(\vec{x},t)=\nabla \Omega (\vec{x},t) }\eqno (9)$$
$$\Omega (\vec{x},t) = 2\nu \bigl [\partial _t \Phi \, +\,
{1\over 2} ({b^2
\over {2\nu }}+ \nabla b)\bigr ]$$
establishes the Girsanov-type martingale connection of the forward
drift
$\vec{b}(\vec{x},t)=2\nu \nabla \Phi (\vec{x},t)$ with the
(Feynman-Kac,
cf. \cite{blanch,olk2}) potential $\Omega (\vec{x},t)$ of the
 chosen external force field.

One of distinctive features of Markovian diffusion processes
with the positive density $\rho (\vec{x},t)$ is that the
notion of the \it backward \rm transition probability density
 $p_*(\vec{y},s,\vec{x},t)$ can be consistently introduced on 
each finite  time interval, say $0\leq s<t\leq T$:
 $${\rho (\vec{x},t) p_*(\vec{y},s,\vec{x},t)=p(\vec{y},s,\vec{x},t)
 \rho (\vec{y},s)} \eqno (10)$$
so that $\int \rho (\vec{y},s)p(\vec{y},s,\vec{x},t)d^3y=
\rho (\vec{x},t)$ 
and $\rho (\vec{y},s)=\int p_*(\vec{y},s,\vec{x},t)
\rho (\vec{x},t)d^3x$.  
This  allows to define the backward derivative of the process in the
conditional mean (cf. \cite{nel1,blanch,vig,olk} for a discussion of
these concepts in case of the most traditional Brownian motion and
Smoluchowski-type diffusion processes)
$$lim_{\triangle t\downarrow 0} \, {1\over {\triangle t}}\bigl
[ \vec{x} - \int p_* (\vec{y},t-\triangle t,\vec{x},t)\vec{y} d^3y
\bigr ]= (D_-\vec{X})(t)=
\vec{b}_*(\vec{X}(t),t)
\eqno (11) $$
$$(D_-f)(\vec{X}(t),t) = (\partial _t + (\vec{b}_* \nabla )- \nu
\triangle )f(\vec{X}(t),t)$$
Accordingly, the backward version  of the dynamical constraint 
imposed on the acceleration field  reads
$${(D^2_-\vec{X})(t) = (D^2_+\vec{X})(t) = \vec{F}(\vec{X}(t),t) }
\eqno (12) $$
where under the gradient-drift field assumption, $curl \, \vec{b}_*=0$ 
we have  explicitly fulfilled  the forced Burgers equation (cf. (1)):
$${\partial _t\vec{b}_* +  (\vec{b}_*\nabla )\vec{b}_* - 
\nu \triangle \vec{b}_* 
= \vec{F}} \eqno (13)$$
where, \cite{nel,zambr,blanch}, in view of
$\vec{b}_*= \vec{b} - 2\nu \nabla ln \rho $, we deal 
with $\vec{F}(\vec{x},t)$ previously introduced in  (9).
A notable consequence of the involved   backward 
It\^{o} calculus  is that the Fokker-Planck equation (6) can be 
transformed to an \it equivalent \rm  form of:
$${\partial _t\rho(\vec{x},t) = - \nu \triangle \rho (\vec{x},t) - 
\nabla [\vec{b}_*(\vec{x},t) \rho (\vec{x},t)]}\eqno (14)$$
with the very same initial (Cauchy) data 
$\rho _0(\vec{x})=\rho (\vec{x},0)$  as in  (6).  

At this point let us recall that  equations (5) and (6) form a natural 
 adjoint  pair of equations that determine the Markovian  
diffusion process in the chosen time interval $[0,T]$.  
Clearly, an adjoint of  (14), reads:
$${\partial _s f(\vec{x},s) = \nu \triangle f(\vec{x},s) - 
[\vec{b}_*(\vec{x},s)\nabla ] f(\vec{x},s)}\eqno (15)$$
where:
$${f(\vec{x},s)=\int p_*(\vec{y},0,\vec{x},s) f(\vec{y},0)d^3y
\enspace  , }\eqno (16)$$
to be compared with (4),(5) and the previously mentioned 
passive scalar dynamics (2), see e.g. also \cite{woy2}.
Here, manifestly, the time evolution of
the backward drift is governed by the Burgers equation, 
and the diffusion equation (15) is correlated (via the
definition (10)) with the probability density evolution rule (14). 

This pair \it only \rm can be consistently utilized if the 
diffusion proces  is to be  driven by forced (or unforced) 
Burgers velocity fields. 

Let us point out that the  study of diffusion in 
the Burgers flow may begin from first solving the Burgers equation
(12) for a chosen external force field, next 
specifying the  probability density 
evolution (14), eventually ending with the corresponding "passive 
contaminant" concentration dynamics (15), (16). All that remains in 
perfect agreement with  the heuristic discussion of the 
concentration dynamics given in   Ref. \cite{monin}, Chap. 7.3. 
where the "backward dispersion" problem 
with "time running backwards" was found necessary to \it predict \rm 
the concentration.

Let us notice that the familiar logarithmic Hopf-Cole transformation, 
\cite{hopf,flem}, of the Burgers equation into the 
generalised diffusion equation (yielding explicit solutions in the 
unforced case) has received a generalisation in the framework of the 
so called Schr\"{o}dinger boundary-data (interpolation) problem, 
\cite{zambr,zambr1,olk2,blanch,olk,olk1}, see also \cite{alb1,freid}.
In particular, in its  recent reformulation,\cite{blanch,olk2}, 
the problem of 
defining a suitable  Markovian diffusion process  was reduced to 
investigating the  adjoint pairs  of parabolic partial 
differential equations, like e.g. (5), (6) or (14), (15). 
In case of gradient drift fields  this amounts to  
checking  (this imposes limitations on the admissible 
force field potential) whether  the Feynman-Kac kernel 
$${k(\vec{y},s,\vec{x},t)=\int exp[-\int_s^tc(
\omega (\tau ),\tau)d\tau ]
d\mu ^{(y,s)}_{(x,t)}(\omega )}\eqno (17) $$
is positive and continuous in the open space-time
area of interest, and whether it gives rise to positive solutions of
the adjoint pair of  generalised heat equations:
$${\partial _tu(\vec{x},t)=\nu \triangle u(\vec{x},t) -
c(\vec{x},t)u(\vec{x},t)}\eqno (18)$$
$$\partial _tv(\vec{x},t)= -\nu \triangle v(\vec{x},t) +
c(\vec{x},t)v(\vec{x},t)$$
where  $c(\vec{x},t)={1\over {2\nu }} \Omega (\vec{x},t)$ follows
from the previous formulas.
In the above, $d\mu ^{(\vec{y},s)}_{(\vec{x},t)}(\omega)$ is the
conditional
Wiener measure over sample paths of the standard Brownian motion.

Solutions of (18), upon suitable normalisation give rise to the
Markovian  diffusion process with the factorised probability density
$\rho (\vec{x},t)=u(\vec{x},t)v(\vec{x},t)$ wich interpolates between 
the boundary density data $\rho (\vec{x},0)$ and 
$\rho (\vec{x},T)$, with the forward and backward 
drifts of the process defined as follows:
$${\vec{b}(\vec{x},t)=2\nu {{\nabla v(\vec{x},t)}
\over {v(\vec{x},t)}}}
\eqno (19)$$
$$\vec{b}_*(\vec{x},t)= - 2\nu {{\nabla u(\vec{x},t)}
\over {u(\vec{x},t)}}$$
in the prescribed time interval $[0,T]$. 
The transition probability density of this process reads:
$${p(\vec{y},s,\vec{x},t)=k(\vec{y},s,\vec{x},t)
{{v(\vec{x},t)}\over {v(\vec{y},s)}}\enspace ,}\eqno (20)$$
Here, neither  $k$, (17), nor $p$, (20) need to be the 
fundamental  solutions  of appropriate parabolic equations, 
see e.g. ref. \cite{olk2} where an issue of  
differentiability is analyzed.

The corresponding (since $\rho (\vec{x},t)$ is given) transition
probability density, (10),  of the backward process has the form:
$${p_*(\vec{y},s,\vec{x},t) = k(\vec{y},s,\vec{x},t)
{{u(\vec{y},s)}\over {u(\vec{x},t)}}\enspace .}\eqno (21)$$
Obviously, \cite{olk2,zambr}, in the time interval $0\leq s<t\leq T$
there holds: \\
${u(\vec{x},t)=\int u_0(\vec{y}) k(\vec{y},s,\vec{x},t) d^3y}
$ and
$v(\vec{y},s)=\int k(\vec{y},s,\vec{x},T) v_T(\vec{x})d^3x$.

By defining $\Phi _*=log\, u$, 
we immediately recover the traditional form of the Hopf-Cole
transformation 
 for Burgers velocity fields: $\vec{b}_*=-2\nu \nabla \Phi _*$.
In the special case of the standard free Brownian motion, there 
holds $\vec{b}(\vec{x},t)=0$ while $\vec{b}_*(\vec{x},t)=-2\nu 
\nabla log\, \rho (\vec{x},t)$.

\vskip0.5cm
{\bf Acknowledgement}: One of  the  authors (P. G. )
receives  a financial support from the KBN research grant
No 2 P302 057 07.

\end{document}